# PERFORMANCE COMPARISON OF PACKET SCHEDULING ALGORITHMS FOR VIDEO TRAFFIC IN LTE CELLULAR NETWORK

Biswapratapsingh Sahoo
Department of Computer Engineering
Indian Institute of Technology, Kanpur, India

bsahoo@iitk.ac.in

## ABSTRACT

*In this paper we have studied downlink packet scheduling algorithms proposed for LTE cellular networks. The study emphasize on three most promising scheduling algorithms such as: FLS, EXP rule and LOG rule. The performance of these three algorithms is conducted over video traffic in a vehicular environment using LTE-Sim simulator. The simulation was setup with varying number of users from 10 - 60 in fixed bounded regions of 1 km radius. The main goal this study is to provide results that will help in the design process of packet scheduler for LTE cellular networks, aiming to get better overall performance users. Simulation results show that, the FLS scheme outperforms in terms of average system throughput, average packet delay, PLR; and with a satisfactory level of fairness index.*

## 1. INTRODUCTION

In the growing complexity of today's wireless communication systems; the cellular data networks are experiencing an increasing demand for its high data rate and wide mobility. In 2008, the Long Term Evolution (LTE) was introduced by the 3rd Generation Partnership Project (3GPP) and in order to increase the capacity and speed of wireless data networks. LTE and WiMAX are the two promising standards in the current cellular technologies, and they are also marketed as 4th Generation (4G) wireless service. LTE offers several important benefits for the subscribers as well as to the service provider. It significantly satisfies the user's requirement by targeting the broadband mobile applications with enhanced mobility. With the introduction of Smartphone, the application like HD TV, online gaming, video meetings, etc are certainly become more valuable to the users. Hence, the users understand and appreciate the benefits of LTE high data-rates and services.

As one of the primary objective of the LTE network is to enhance the data-rate so as to cater the range of highly demanded services; the radio resources or data channels are divided and shared efficiently among different active users while considering a satisfied level of QoS to all active users. To shape the requirements, the LTE system uses orthogonal frequency division multiple access (OFDMA) technology in the downlink. The OFDMA technology divides the available bandwidth into multiple narrow-band sub-carriers and allocates a group of sub-carriers to a user based on its requirements, current system load and system configuration [1]. On that ground, collisions can happen more often among the active users while sharing the same data channel. Hence it is need to be taken care by the resource allocation scheme.

With 3G technologies, simple video services were attempted with attention focused on video telephony and multimedia messaging; and were somewhat successful. However, these technologies are still capacity limited and not able to economically support the huge mobile video demand that is emerging. This directly leads to the desire to use a broadband wireless network, such as LTE, for doing the same stuffs with greater speed and mobility. This can be mostly achievable by efficient resource allocation technique. Hence the fast packet scheduling is one of the important data-rate improvement techniques among Link Adaptation, Multiple-Input and Multiple-Output (MIMO) Beam forming and Hybrid ARQ in 4G networks.

In order to support the rich demand of real-time multimedia services like video streaming, VoIP, and etc., it is necessary to ensure the quality-of-service (QoS) requirements are met and packet loss ratio (PLR) is minimised by keeping it below the application's required threshold. In a video streaming service environment, it is important to maintain the PLR threshold below 1% [2] such that the QoS requirements of video streaming service users are satisfied.

In this paper, we have studied packet scheduling algorithms that are proposed in past year for LTE cellular networks implication to video traffic. A comparison of performance indexes such as: average system throughput, average delay, PLR and fairness is reported for LTE systems over a realistic simulated scenarios in multi-cell environment. The main goal this study is to provide results that will help in the design process of LTE cellular networks, aiming to get better overall performance.

The rest of the paper is organized as follows. An analysis of packet scheduler proposed in past year for LTE systems over video traffic is presented in Section II. Section III describes the fast packet scheduling algorithms which are used in the simulations. Section IV describes simulation conditions, scenario schemes and traffic model. Simulation results are presented in section V. Finally, Section VI concludes the paper.

## 2. BACKGROUND STUDIES

Scheduling is one of the main features in LTE systems; because it is in charge of distributing available resources among the active users in order to satisfy their QoS needs [3]. In LTE network architecture, packet schedulers (for both the downlink and the uplink) are deployed at the eNodeB (eNB). The packet scheduler plays a fundamental role: it aims to maximize the spectral efficiency through an effective resource allocation policy that reduces or makes negligible impact of the channel quality drops. Now-a-days the real-time multimedia services such as video and VoIP are becoming increasingly popular among the users in recent cellular technologies. The LTE provides, QoS of multimedia services with fast connectivity and high mobility. However, 3GPP specifications have not defined any scheduling policies to support real-time or non-real time application services [4], but leaves its design and implementation to the vendors. A generalized working model of packet scheduler for downlink LTE system is shown in Fig. 1.

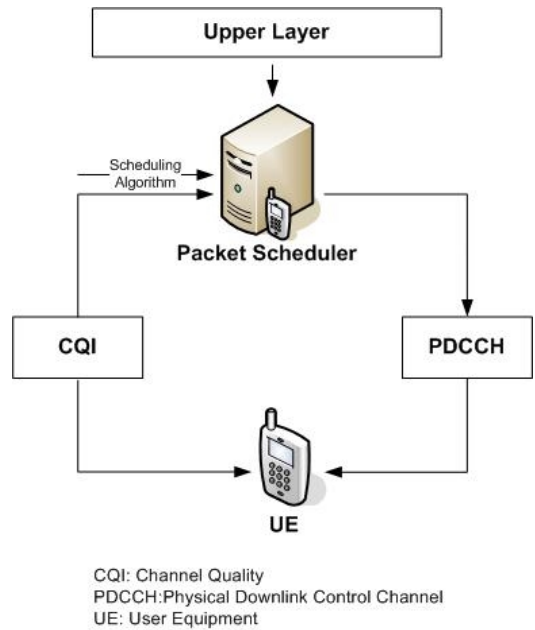

Figure 1. Working Model of Packet Scheduler for Downlink LTE System

To satisfy the QoS requirements of video users, the scheduler that aim guarantee bounded delay, are to be the suitable schemes; since one of the main requirement of video streaming is, to delivered the data packet within a certain deadline. In general, the packet scheduler prioritizes the users, based on certain criteria such as channel condition, packet delay and type of service or traffic. In addition to these parameters, [5, 6, 7] has taken content of the video streaming into consideration for prioritizing the users turn or data packet. The authors in [8] proposed a genetic algorithm based approach to optimize the application-layer video quality. They have used the rate-distortion theory based method to estimate the video quality and exploit genetic algorithm based metaheuristic approach to improve the accuracy of prioritization.

By controlling the transmission of enhancement layer information in each scheduling round, the performances can be improve in terms of the packet delivery ratio and also it is possible to maintain better service qualities [9]. The author in [9] has considered scalable video coding (SVC) into account. Supporting the performance of the work of [9], the study in [10] claims that the scheduling schemes based on SVC performs better with acceptable video quality in comparison to H.264.

To satisfy the QoS requirements of video users, the researchers always faces the trade-off between decision optimality and computational complexity while designing a resource allocation strategy. The study suggests, while defining a resource allocation policy the LTE vendor should take the design factors into account such as [3]: complexity and scalability; spectral efficiency; fairness and QoS provisioning.

This paper evaluates three most promising channel-aware/QoS-aware strategies (i.e., Frame Level Scheduler (FLS) [11], Exponential (EXP) rule [12] and Logarithmic (LOG) rule [13]) in a realistic scenario over LTE network. The performance result is presented in Section V.

## 3. PACKET SCHEDULING SCHEMES

The purpose of Packet Scheduling is to distribute resources among users in a fair and efficient way to maximize the system throughput along with fairness. The scheduler is one of the key elements and to a large extent determines the overall downlink performance, especially in a highly loaded network. In this section, we will illustrate the working principle of three different packet scheduling algorithms for LTE: FLS, EXP rule and LOG rule.

The FLS [11] is a two-level scheduling scheme with one upper level and lower level. Two different algorithms are implemented in these two levels. A low complexity resource allocation algorithm based on discrete time linear control theory is implemented in the upper level. It computes the amount of data that each real-time source should transmit within a single frame, to satisfy its delay constraint. The proportional fair (PF) scheduler, originally proposed in [14], is implemented in the lower level to assign radio resources to the user, to ensure a good level of fairness among multimedia flows. The following equation calculates the amount of data to be transmitted.

$$u_i(k) = h_i(k) * q_i(k)$$

where $u_i(k)$ corresponds to the amount of data that is transmitted during the $k^{th}$ frame; '*' operator is the discrete time convolution. The above equation tells that the amount of data to be transmitted by the $i^{th}$ flow during the $k^{th}$ LTE frame is obtained by filtering the signal $q_i(k)$ (i.e., the queue level) through a time-invariant linear filter with pulse response $h_i(k)$.

Moreover, according to the results presented in [11], the FLS is a good candidate for guaranteeing bounded delays to multimedia flows, and also provides the lowest PLR. Thus, it ensures the highest video quality to mobile users.

The EXP rule [12] was proposed to provide QoS guarantees over a shared wireless link. While scheduling the active users in the network, a wireless channel is shared among multiple numbers of users and each user's data arrives to a queue as a random stream where it awaits transmission/service. A scheduling rule in this context selects a single user/queue to receive service in every scheduling instant. In [12], author proposed a new scheduling policy and proved that the proposed rule is throughput-optimal. This scheduling scheme explicitly uses the information on the state of the channel and queues and it ensures the queues' stability without any prior knowledge of arrival and channel statistics of traffic.

The EXP rule scheme defined two exponential rule for service a queue: the EXP (Queue length) rule (EXP-Q) and the EXP (Waiting time) rule (EXP-W). It chooses either EXP-Q or EXP-W rule for a service a queue with fixed set of positive parameters such as: $\beta, \eta \in (0, 1)$ and $\gamma_i, a_i, i \in N$.

$$i \in i(S(t)) = \arg\max_i \gamma_i \mu_i(t) \exp\left(\frac{a_i Q_i(t)}{\beta + [\overline{Q}(t)]^\eta}\right)$$

$$i \in i(S(t)) = \arg\max_i \gamma_i \mu_i(t) \exp\left(\frac{a_i W_i(t)}{\beta + [\overline{W}(t)]^\eta}\right)$$

where $\mu_i(t)$ is the number of user served from the queue at time t, $\overline{Q_i} \doteq (\frac{1}{N})\sum_i a_i Q_i(t)$ and $\overline{W_i} \doteq (\frac{1}{N})\sum_i a_i W_i(t)$.

To provide a balance between mean-delay and throughput, the EXP rule maintains a lower delay by compromising the throughput, and eventually the mean delays and tails for almost all users [15].

The LOG rule [13] scheduler was design to provide a balance in QoS metrics in terms of mean delay and robustness. It allocates service to a user in the same manner as EXP rule to maximize the current system throughput, assuming the traffic arrival and channel statistics are known. The simulations in [13] for a realistic wireless channels shows the superiority of the LOG rule which achieves a 20-75% reduction in mean packet delay. The author claims that the LOG rule is for a practical solution but, is not provably mean-delay optimal.

The wireless channel is time-varying; hence the transmission rates supported for each user vary randomly over time. If the channel state is available in prior, a prioritizing policy can be implemented to schedule the active users so as to exploit favourable channels, e.g., schedule the user which currently has the highest rate – this is referred to as opportunistic scheduling [13]. The two discussed algorithm (EXP rule and LOG rule) are referred in the category of opportunistic scheduling. A comparison study of the above two schemes LOG rule and EXP rule is presented in [15]. Furthermore, EXP and LOG rules have been proved as the most promising approaches for downlink scheduling in LTE systems with delay-sensitive real-time applications like video and VoIP.

## 4. SIMULATION FRAMEWORK

In this section, we will discuss the simulation setup which is used in this paper to evaluate the three different scheduling schemes in video traffic scenarios over LTE networks. To compare the performance of the above discussed scheduling algorithms, the LTE-Sim [16], an open source simulator for LTE networks was used. The results of the simulation are discussed in next section.

### 4.1. Simulation Scenario

To perform the test, we defined a multi-cell scenario with a fixed eNodeB at the centre of the cell. The users are moving in a vehicular propagation environment with a speed of 120 kmph. Mobility of each user equipment (UE) travelling cells is described with the random direction model. The users are moving in a bounded region of radius equal to 1 km. The number of connections to an eNodeB at a particular time varies from 10 - 60 UEs. Each UE receives one video flow, one VoIP flow, and one best-effort flow at the same time. Table 1 gives the details simulation parameters used in the created simulation scenario for this paper.

Table 1.  Simulation Parameters

| Parameters | Value |
| --- | --- |
| Simulation Duration | 100 Sec |
| Frame Structure | FDD |
| Radius | 1 KM |
| Bandwidth | 10 MHz |
| Slot Duration | 0.5 ms |
| Scheduling Time (TTI) | 1 ms |
| Sub-career Spacing | 15 kHz |

International Journal of Mobile Network Communications & Telematics ( IJMNCT) Vol. 3, No.3, June 2013

| Maximum Delay | 0.1 ms |
|---|---|
| Video bit-rate | 242 kbps |
| User Speed | 120 km |
| Traffic Model | H.264, Random Direction |

### 4.2. Video Traffic Model

The video traffic is treated as a real-time multimedia service. The real time video streaming traffic model can be described as continuous occurrence of video-frames in a specific time interval. Video traces can be employed to simulate video traffic in a wide range of network simulations [17]. Video traces are used to generate video traffic workloads as well as to estimate the video related performance metrics, such as the frame starvation probability. In the studies reported here, in order to obtain a realistic simulation of an H.264 SVC video streaming, we have used a encoded video sequence i.e. ``foreman.yuv``. The encoded spatial resolution CIF 352x288 with 300 frames per second has been used for the entire simulation.

## 5. SIMULATION RESULTS AND DISCUSSIONS

In this section, the performance result of three different schemes such as FLS, EXP rule and LOG rule have been analyzed and evaluated by varying the number of UEs while considering the inter-cell interference. Each simulation last upto 100 sec in a realistic multi-cell scenario and all simulation results are averaged over five simulations. The comparison has been divided on the basis of several performance indexes of each algorithm such as the PLR, average system throughput, average delay, fairness and spectral efficiency.

The system throughput with increasing number of users is shown in Figure 2. It is observed that the throughput degrades with increasing number of users and it is demonstrated within the figure that the FLS is having the best throughput performance followed by the EXP-rule and LOG-rule algorithms.

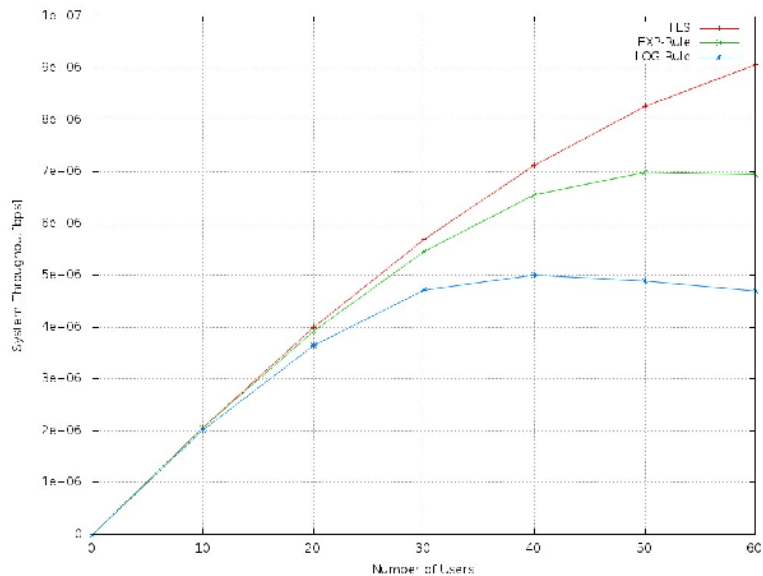

Figure 2. System Throughput vs. Number of Users

The FLS; since in upper level it estimates the amount of data that should be transmitted in a single time frame, it favours to degrade the PLR and the effect hence be resulted in system throughput gain.

Figure 3 shows the packet delay performance under the same three algorithms. It can be notice from the figure that, the FLS scheme is having better performance than EXP-rule and LOG-rule schemes. On the other hand, both EXP-rule and LOG-rule schemes has a similar delay on a higher number of users. Hence it can be conclude from the result that, FLS supports best in terms of packet delay over video traffic. The import ants metrics here is, as the real-time applications are delay sensitive, we can consider the FLS as the best suited schemes for real-time applications out of these three discussed schedulers.

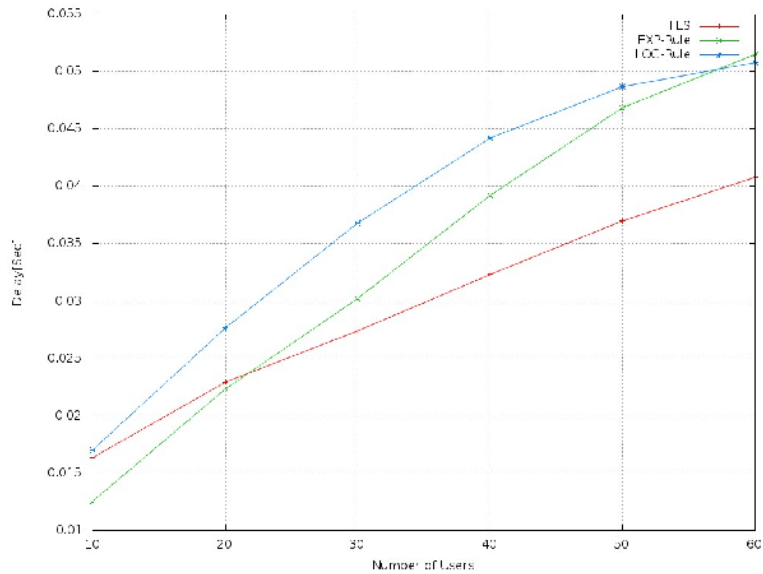

Figure 3. Delay vs. Number of Users

The impacts of varying PLR, while the number of users is increasing are shown in the Fig. 4. A significant difference can be found between these three schemes in PLR performance. The LOG rule has the highest PLR with higher number of subscribers followed by EXP-rule. As cited previously, it is important to maintain the PLR threshold below 1% for video streaming; the performances of EXP-rule and LOG-rule are acceptable but they may not be the suitable schemes for critical video applications or real-time applications. In such case we need to choose a scheme having comparatively lower PLR.

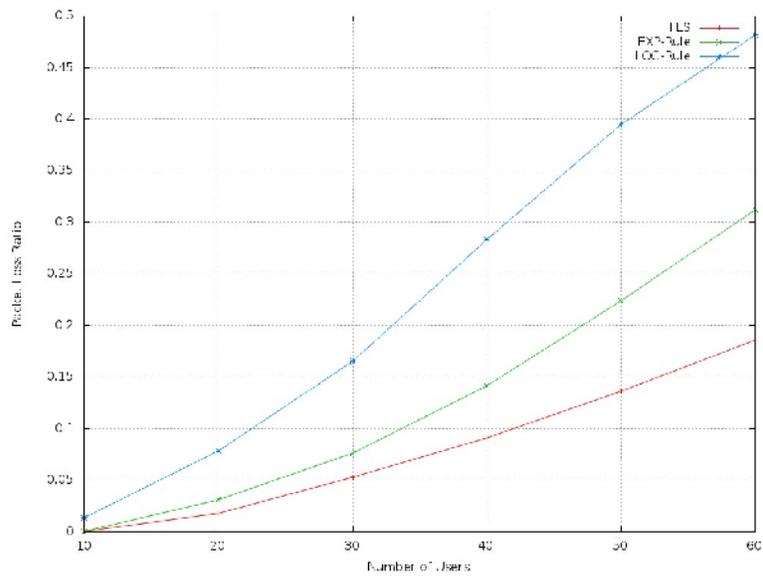

Figure 4. Packet Loss Ratio vs. Number of Users

The fairness performance of each packet scheduling algorithm is shown in Fig. 5. It can be detected that the three algorithms are in a race with each other while number of users varies. Since there is no significant difference in the simulation results of these algorithms, it is still unsure that which algorithm is best suited for video streaming.

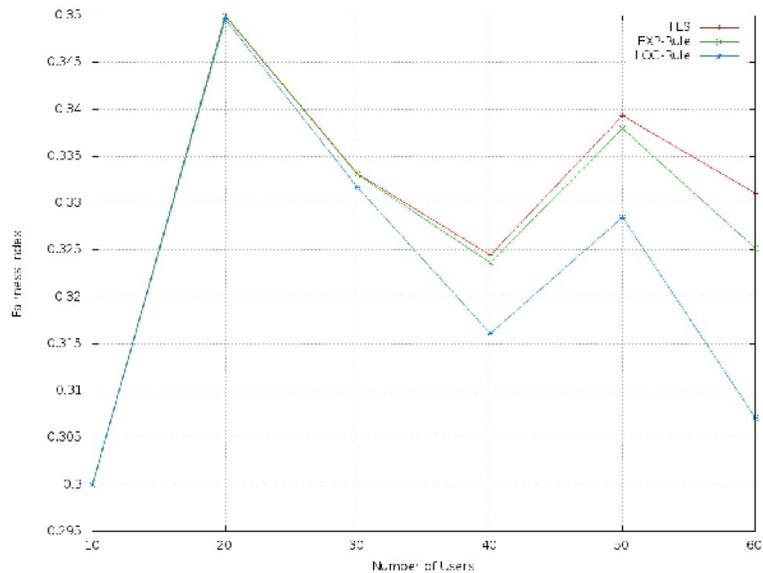

Figure 5. Fairness vs. Number of Users

Finally, in Fig. 6, we verified the spectral efficiency over the varying number of subscribers within the scenario. With increasing the number of subscribers, they do not have any different impact towards the efficiency performance. Since they perform almost with similar measure with varying number of users, which algorithm can best support video streaming services is inconclusive.

16

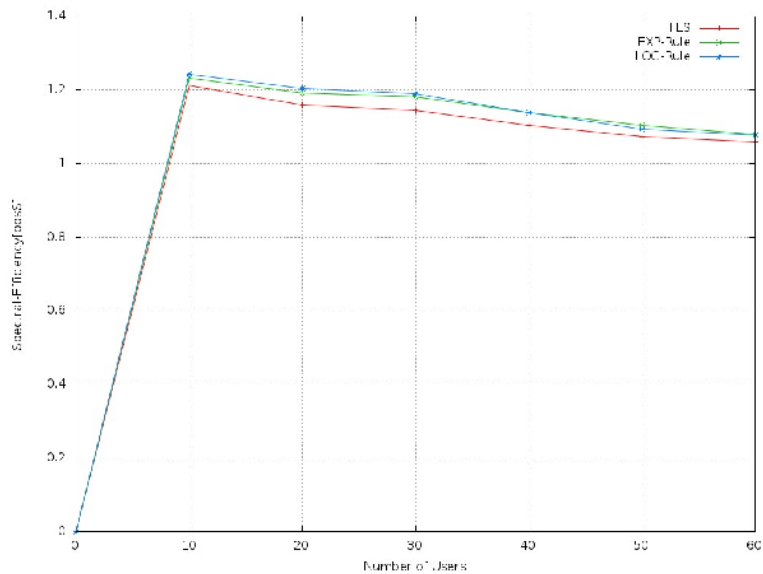

Figure 6. Spectral efficiency vs. Number of Users

## 6. CONCLUSIONS

This paper evaluates the performance of three promising downlink packet scheduling algorithms for LTE cellular networks. The study was analyzed with performance matrices such as average system throughput, average packet delay, PLR, fairness and spectral efficiency over video traffic in a vehicular environment. The simulation results show that, the FLS algorithm outperforms over other schedulers in terms of higher system throughput, low packet delay, low PLR and satisfactory fairness index. Hence it is conclude that FLS is the best candidate for downlink packet scheduling algorithm LTE cellular networks over video traffic in vehicular environment.

## ACKNOWLEDGEMENTS

The author is thankful to Dr. Kumbesan Sandrasegaran, Professor, Faculty of Engineering & Information Technology, University of Technology, Sydney, Australia for his kind support.

**Authors**

Biswapratapsingh Sahoo is presently working as a Project Engineer at Indian Institute of Technology , Kanpur India. He holds a Master Degree in Computer Science from Utkal University, India and a Bachelor of Science Degree in Computer Science from Ravenshaw University, India. He is broadly interested in the areas of wireless network and mobile communication systems.

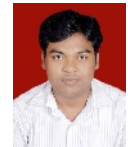